# Extremely efficient clocked electron transfer on superfluid helium

F. R. Bradbury,[1,†] Maika Takita,[1] T. M. Gurrieri,[2] K. J. Wilkel,[2] Kevin Eng,[2,‡] M. S. Carroll,[2] and S. A. Lyon[1]

[1]Department of Electrical Engineering, Princeton University, Princeton, New Jersey, USA
[2]Center for Integrated Nanotechnologies, Sandia National Laboratories, Albuquerque, New Mexico, USA

[†]Present address: Amsterdam University College, Amsterdam, The Netherlands
[‡]Present address: HRL Laboratories, LLC, Malibu, CA

**Unprecedented transport efficiency is demonstrated for electrons on the surface of micron-scale superfluid helium filled channels by co-opting silicon processing technology to construct the equivalent of a charge-coupled device (CCD). Strong fringing fields lead to undetectably rare transfer failures after over a billion cycles in two dimensions. This extremely efficient transport is measured in 120 channels simultaneously with packets of up to 20 electrons, and down to singly occupied pixels. These results point the way towards the large scale transport of either computational qubits or electron spin qubits used for communications in a hybrid qubit system.**

Interesting applications of quantum algorithms will need large numbers of logical qubits, as well as the many-fold redundancy of corresponding physical qubits required for quantum error correction.[1] While experiments have shown several simple quantum systems to have qubit coherence times exceeding the time necessary for performing operations,[2-6] thus far none have demonstrated scaling beyond a handful of coupled qubits.

Electrons floating on the surface of liquid helium experience a uniquely small spin-orbit interaction for a condensed matter system, and are expected to have long spin coherence times.[7] Here we demonstrate exceptionally efficient clocked electron transport along gate-defined paths across the surface of superfluid helium. The gates form the equivalent of a buried-channel charge-coupled device (CCD),[8] which simultaneously moves electron packets in many parallel channels (120 channels in the structure used for this work). No transfer errors were detected after clocking the electron packets across $10^9$ pixels (moving them 9 km), even with one electron per packet, on average. This highly parallel and efficient electron transport could be leveraged in a hybrid quantum processor to form a quantum communications fabric between computational qubits defined and localized in the substrate, for example superconducting qubits.[9] Alternatively, the same electron spin qubits could be used for both communications and computation.[7]

Electrons are held above the surface of superfluid helium by the fields from underlying electrostatic gates (in addition to a weak image potential) and their positions can be controlled by voltages applied to those gates.[10-13] They form a clean and well-studied classical two dimensional electron system (2DES). Besides being the highest mobility 2DES[14] and forming the first demonstrated 2D Wigner crystal,[15] their transport properties have been used to probe quantized excitations of both the electrons' bound states[16] and the underlying superfluid.[17] Our work takes inspiration from previous studies which utilized lithographically-defined channels that fill with superfluid helium by capillary action for a well-defined helium surface geometry.[18-23]

The helium channel CCD devices were fabricated using a complementary metal-oxide-silicon (CMOS) process at Sandia National Laboratory.[24] The top two layers of metal interconnects are used to define the channels and gate electrodes, while underlying metal planes shield the modulation and electron sensing structures from carriers in the silicon substrate. The devices must undergo post-processing reactive ion etching to remove $Si_3N_4$ and $SiO_2$ above the relevant gates. They are wired and mounted in a cell that is cooled to 1.6K. The chip is positioned channel-side up about 1 cm above the bulk surface level of the superfluid helium. The channels fill with helium by capillary action. Electrons are photoemitted from a zinc film into the vacuum above the device[25] and attracted to the channels by a positive bias on the large right or left *reservoir* gates. Figure 1 illustrates the layout and functionality of the device's array of gates and channels.

Electrons are detected non-destructively through their capacitive coupling to the *sense* gate (a method similar to the floating-gate amplifiers employed in early Si-CCDs[26]). The electrons' coupling to the sense gate is modulated as they are pushed over it and pulled away by a constant ac-voltage ($V_{ac}$) applied to the neighbouring *twiddle* gate of 400 mV$_{pp}$. The *door* gate's dc-voltage ($V_{dc}$) controls the loading of electrons from the right reservoir, while the amplitude and phase of its $V_{ac}$ are adjusted to cancel the direct capacitive coupling between the twiddle and sense gates. During a measurement, the door gate is "closed" with a $V_{dc}$ which is negative (repulsive) with respect to the twiddle and sense gates. The oscillating motion of the electrons induces a periodic voltage on the sense gate at the frequency of the twiddle gate's $V_{ac}$. This induced voltage is buffered by a high electron mobility transistor (HEMT) located on the backside of the printed circuit board to which the device is mounted. Due to the small gates and narrow channel geometry, the differential voltage induced on the sense gate by one electron corresponds to only a fractional charge. The capacitance calibration relating signal magnitude to electron number can be calculated and approximately confirmed by measuring the signal's non-linear roll-off as the electron density is increased to saturation (for a particular ac driving voltage). Noise levels for the electron measurement are found to be ~360 e/√Hz, or ~3 e/channel/√Hz, since electrons in all 120 channels are measured together.[27]

A typical experiment begins with emptying all the CCD pixels, then loading electrons from the right reservoir onto the twiddle and sense gates. The desired pixel occupancy is obtained by temporarily lowering the potential of the door gate to bleed off excess electrons back to the reservoir. After the initial charge measurement, this packet of electrons is loaded into one pixel of the gates connected as CCDs and clocked along the channels using a programmed voltage sequence on the CCD gates. Electrons can be moved back to the detection gates for a charge measurement after any number of clock cycles in the CCDs.

The first test of these devices used the horizontal 3-phase CCD gates to move electrons back and forth in the 120 parallel channels. As seen in Fig.1, electrons moving along these channels must traverse 3- and 4-way channel intersections and a constriction of channel width. These irregularities change the electric potential along the channels. However, we find that when the CCD gate excursions (clocking voltages) are increased to 3V, electrons can be transferred efficiently across all gates in the parallel channels, even with a packet of 20 or more electrons loaded onto a pixel.

Figure 2 presents results from an experiment to test electron transfer efficiency in which one packet (per channel) of electrons is moved to the rightmost pixel of the 3-phase CCD structure, and then clocked 10 pixels to the left and 10 back to the right along the horizontal channels. After repeating this left-right clocking a number of times the electron packet is moved back to the detection gates for measurement. If the electrons transfer reliably, then they all return to the same pixel from which they began the back and forth sequence. We use the detection gates to measure electrons on the correct pixels as well as those ending up on pixels to the right and left. The graph plots the signal from electrons which were returned to the correct pixel after clocking them for a given number of cycles at a pixel-to-pixel clocking rate of 240 kHz. At these clocking speeds, there is no measurable signal loss, meaning that all the electrons are returning to the correct gate after over a billion pixel cycles (over an hour of clocking). No electrons are measured on the adjacent "wrong" pixels. Further experiments to explore the limits of reliable clocking show that the clocking rate is limited by the RC rise time of the cryostat wiring in these experiments.[27] No measurable loss of electrons is found for packets containing several electrons (blue data set), nor for packets containing less than one electron on average (red data set).

The upper bound of the transfer failure rate is at least four orders of magnitude lower than in semiconductor CCDs.[28] This extraordinary improvement is expected due to the high electron mobility and strong fringing fields across the CCD gates. Finite element modelling shows that fringing fields above gates and in the direction of the channels are ~700 V/cm for the clock voltages used in these experiments. Electrons subject to such fields will traverse a 3 μm gate in ~20 ps. While these gate sizes would be small enough to observe hot electron effects, the temporal control is not sufficient as gate voltage rise times are ~100 ns. Therefore, the extremely high transfer efficiency (at our RC-limited clocking frequencies) establishes that the disorder potential at the helium surface is small compared to the gate fringing fields.

In addition to the 120 parallel channels, our device also has a single perpendicular channel (vertical on the micrograph of Fig.1) with underlying CCD gates. This perpendicular CCD is used to demonstrate 2D transport and also measure population uniformity across the 120 parallel channels. To measure population uniformity, electrons are clocked up or down the vertical channel, then brought back to the detection gates for a charge measurement. The electron signals decrease due to this upward or downward clocking because electrons are progressively ejected from the topmost or bottommost channel. The ejected electrons likely find their way onto the thin van der Waals helium film covering the top metal plane and are immobilized by a spatially varying image potential due to the metal's roughness. Experimentally, we find that ejected do not re-enter the channels.

Details of channel occupancy experiments are described elsewhere.[27,29] We find that the populations of all 120 channels can be equalized by filling the detection gates with large electron packets; then controllably bleeding electrons away to the desired occupancy by lowering the potential of the door gate. This method of obtaining approximately uniform channel populations was employed for all of the reported electron transfer efficiency experiments.

Successful transport along the vertical channel has important implications for flexible and scalable qubit movement and control. To test the reliability of 2D transport in the channel array, we created the clocking sequence sketched in the inset of Fig.3. The electrons are moved in a "C" pattern: one pixel left, 60 pixels up, one pixel right, and then the sequence is reversed to return the electron packets to their starting point. Larger clocking voltage excursions (5V) are necessary for reliable transfer along the narrower vertical channel since the gates are screened more than in the wider channels. With 5V clock voltages we find that the electrons can be clocked in this C-pattern for over a billion pixel cycles at the same 240 kHz pixel frequency with no measurable failures, as shown in Fig.3.

Our demonstration of efficient electron transport on helium shows how an electron spin array can be scaled to allow complex and parallel operations on many qubits. Other experiments with this device demonstrate the ability to load and store electrons in the memory cells and to unload them as needed at a later time. The immeasurably low transfer failure rate after more than a billion pixel cycles through constrictions and intersections in channels of differing widths proves the flexibility and 2D scalability of this approach. Data from the C-pattern sequence demonstrate that transporting electrons around corners incurs no time delay penalty nor requires special treatment besides the larger clocking voltage used to move electrons along the narrower vertical channel (and these voltages could be reduced by optimizing the channel widths). Five control lines are sufficient to move an electron to any of the 4560 distinct sites of gates and channels using 3–phase CCD clock sequences. Measurements conducted with less than one electron on average per channel demonstrate this extremely high transfer fidelity in the quantized regime.

We would like to thank S. Shankar, G. Sabouret, and D.I. Schuster for helpful discussions and J. Donnal for the electronics to apply clock voltages to the CCD. Work at Princeton was supported by the NSF under grant CCF-0726490. Additionally, this work was performed, in part, at the Center for Integrated Nanotechnologies, a U.S. Department of Energy, Office of Basic Energy Sciences user facility. Sandia is a multiprogram laboratory operated by Sandia Corporation, a Lockheed Martin Co., for the United States Department of Energy under Contract No. DE-AC04-94AL85000.

Correspondence should be addressed to S.A.L. (lyon@princeton.edu).

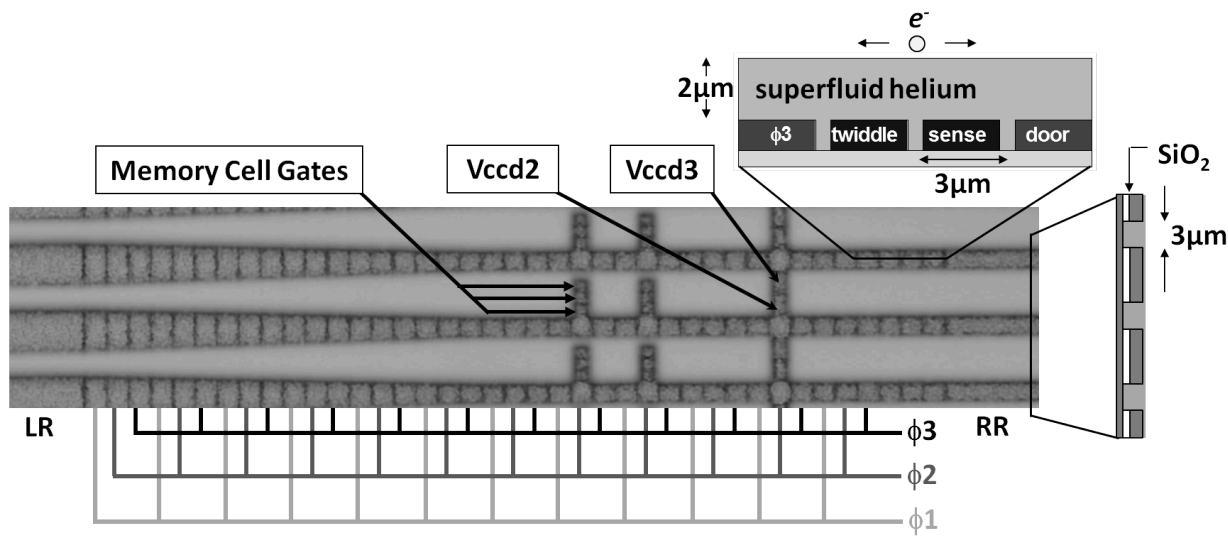

Figure 1. Micrograph showing a small central portion of the helium channel CCD device where the electron clocking experiments are performed. Three of the 120 parallel horizontal channels are pictured, with a schematic cross-section to the right of the image. The darker parts of the image are the channels with underlying gates, and the lighter parts are the topmost metal layer ridges between the channels. Electrons are confined above the helium-filled channels by a positive dc bias (3V) on the gates with respect to the top metal plane. Just visible on either side of the image are the left and right electron collection reservoirs, labelled "LR" and "RR", with channel widths of 6 μm and 3 μm, respectively. The small gates in the horizontal channels between the right and left reservoirs have a 3 μm period including a 0.5 μm gap and are arranged as a 3-phase CCD, connected by on-chip vias to clock rails in a lower metal layer (drawn schematically below the micrographs). Each set of three CCD gates makes up one pixel for electron control. Exceptions to the 3-phase connections are the three gates closest to the right reservoir, which are labelled "twiddle", "sense", and "door" and serve as loading and detection gates. The cartoon image shows an electron on the helium surface in a cut-away view along the middle of a channel. The horizontal channels are interposed by two memory cells for electron storage and an interconnecting vertical CCD channel (clocked with voltages on φ2, Vccd2, and Vccd3) for inter-channel transport which are all 2.5 μm wide. All gates used in these experiments run under and control electrons in all 120 channels in parallel.

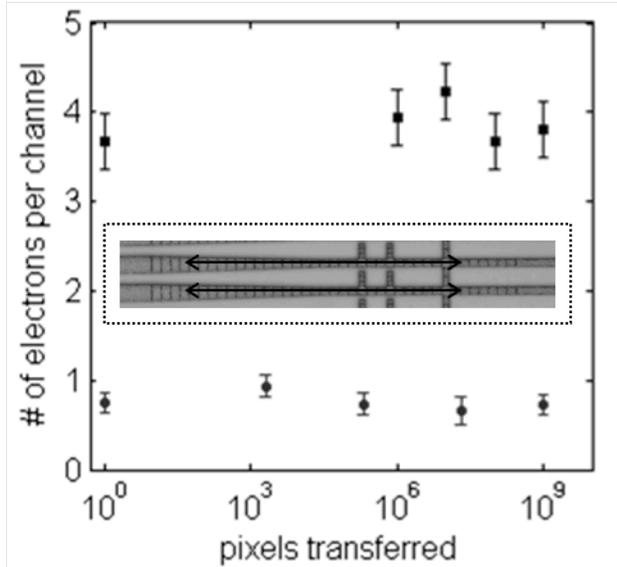

Figure 2. Measurement of the electron transfer efficiency along horizontal channels. The superimposed arrows in the inset image show the path electrons are clocked back and forth in the channels. The graph shows the electron signal measured at the correct pixel after a given number of CCD cycles, demonstrating an immeasurably low occurrence of transfer failure. The blue data were obtained in an experiment with about four electrons per packet (averaged over the 120 parallel channels), while the average packet size was less than one electron (~0.75) for the experiment represented by the red data.

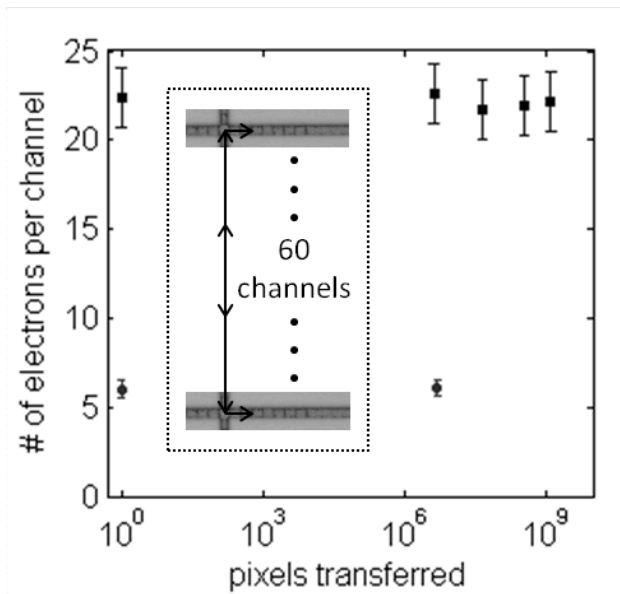

Figure 3. Measurement of the electron transfer efficiency in 2D. The inset shows how electrons are cycled in a "C" pattern in the horizontal and vertical channels. The graph plots the measured signal from electrons which return to the correct pixels after a given number of cycles. The channel occupancy is calculated by dividing the calibrated signal by 60 (instead of 120 channels) because electrons in the upper 60 channels are ejected during the first instance of the C-pattern. Two data sets are shown depicting experiments performed with different initial pixel populations.